\documentclass[useAMS,usenatbib]{mn2e}
\input {psfig.sty}
\usepackage{amsmath}
\usepackage{amssymb}
\def \bh {_{\rm BH}}
\def \blr {_{\rm BLR}}
\def \Ms {$M-\sigma$}

\def \mgii {Mg\,{\sc ii}}
\def \civ {C\,{\sc iv}}
\def \hbeta {H$\beta$}
\def \oiii {[O\,{\sc iii}]}
\def \oii {[O\,{\sc ii}]}

\def \empha {({\emph a})}
\def \emphb {({\emph b})}

\title[Orientation effects in quasar spectra]{Orientation effects in quasar spectra: The broad- and narrow-line regions}

\author[Fine, Jarvis \& Mauch]
       {S. Fine$^1$\thanks{stephen.fine@durham.ac.uk},
	 M.~J. Jarvis$^2$, T. Mauch$^3$   \\
$^1$Department of Physics, Durham University, South Road, Durham DH1
	 3LE, UK \\
$^2$Centre for Astrophysics Research, Science \&\ Technology Research Institute,
       University of Hertfordshire, Hatfield AL10 9AB, UK \\
$^3$Oxford Astrophysics, Denys Wilkinson Building, Keble Rd, Oxford
       OX1 3RH, UK \\
}

\begin{document}

\maketitle

\begin{abstract}

We use the Sloan Digital Sky Survey, along with the NRAO VLA Sky
Survey and the Westerbork Northern Sky Survey to define a sample of
746 radio-loud quasars 
and measure their 330\,MHz to 1.4\,GHz spectral indexes. Following
previous authors we take the
radio spectral index as an indicator of the orientation towards the
quasars such that more pole-on sources tend to have flatter spectral indexes.
We use this proxy for the orientation of quasars to investigate the
effect orientation may have on optical spectra.

Quasars with flatter spectral indices tend to be
brighter. However, we find no indication of reddening in
steep-spectrum QSOs to indicate obscuration of the accretion
disk by a torus as a possible explanation. Nor do we find increased
redddening in the flat-spectrum sources which
could imply a contribution from jet-related synchrotron emission. 

We reproduce a previously-described anti-correlation between the width
of the \mgii\ line and radio
spectral index that indicates a disk-like geometry for the \mgii\
BLR. However, in contrast to previous authors we find no such
correlation for the \civ\ line suggesting a more isotropic
high-ionisation BLR.

Both the \oii\ and \oiii\ narrow lines have more
flux in steep spectrum sources while the \oiii/\oii\ flux ratio is
lower in these sources. To describe both of these effects we propose a
simple geometric model in which the NLR exists primarily on the surface
of optically thick clouds facing the active nucleus and the NLR is
stratified such that higher-ionisation lines are found preferentially
closer to the nucleus.

Quantitatively we find that orientation may effect the observed strength
of narrow lines, as well as ratios between lines, by a factor of $\sim$2.
These findings have implications for the use of \oiii\ and \oii\
emission lines to
estimate bolometric luminosities, as well as comparisons between
narrow line luminosity functions for type~1 and type~2 objects and the
potential of emission-line diagnostic diagrams as an accurate tool
with which to distinguish types of active galactic nuclei. 

Finally we find no evidence that BAL
QSOs have a different spectral index distribution to non-BALS although
we only have 25 obvious BALs in our sample.

\end{abstract}


\section{Introduction}

Orientation plays a key role in the unification model for active
galactic nuclei (AGN;
\citealt{ant93}) as one of the most important parameters that define
the optical
characteristics of an AGN. While we have become
familiar with appealing to orientation in order to explain the
different classes of observed AGN, the effects of orientation
within a single class of objects are less well understood. This paper aims to
study the effect of orientation on spectra of quasi-stellar objects
(QSOs) by using the radio spectral index as a proxy for
orientation. Our findings
can broadly be categorised into those pertaining to the broad-line
region (BLR) and those pertaining to the narrow-line region (NLR). We
will discuss these separately throughout much of the paper. We also
touch on the effect of orientation on the optical continuum shape and
proportion of broad-absorption line (BAL) systems in QSO spectra
later in the paper.


\subsection{Orientation and the BLR}


The BLR is too small to be spatially resolved with current
telescopes, hence the geometry of the BLR, and the physics that
governs its dynamics, is not well constrained. There are
some indications that the dynamics of the BLR may be dominated by the
local gravitational potential (e.g. \citealt{p+w99}). Virial
supermassive black-hole (SMBH) mass 
estimators make use of this assumption to estimate the masses of SMBHs in QSOs.

Integral to SMBH mass estimators for QSOs is an assumption about the geometry
of the BLR. This is commonly expressed as a single factor, $f$, which
translates the measured FWHM of a spectral line into the
velocity of the BLR, $v\blr$,
\begin{equation}
v\blr=\sqrt{f}\times FWHM.
\end{equation}
Hence, given the radius of the BLR ($r\blr$), and assuming that the
local gravitational field governs the dynamics, the SMBH mass ($M\blr$)
can be estimated as
\begin{equation}
M\bh=f\frac{r\blr FWHM^2}{G}
\label{equ:vir_f}
\end{equation}
The value of $f$ is unknown. Model values give $f=3/4$ for a
spherically symmetric BLR \citep{p+w99}, or $f=1/(4\sin^2\theta)$ for a
disk BLR inclined at an angle $\theta$ to the observer \citep{m+d02},
hence orientation can affect the value of $f$ and can bias SMBH mass
estimates \citep{j+m02}.

Given the interest in BLR dynamics, and their importance to SMBH mass
estimation, there have been many studies of the structure of the
BLR. Techniques employed include
defining a proxy for orientation and using it to
look at the effect on broad lines \citep{bro96,j+m06},
working out SMBH masses through a method independent of the virial
technique and then working backwards through the virial equation to
give $f$ \citep{m+d02,onk04,lab06},
looking at the broad line distribution as a whole and
gleaning results from that \citep{ost77,me2}, velocity-resolved
reverberation mapping \citep{koll96,den10,ben10} and recently through
X-ray absorption \citep{ris10}.

In radio-loud QSOs the ratio ($R$) between core and lobe emission has been
employed as an orientation indicator under the assumption that core
emission is Doppler-boosted when observed down the axis of the object
\citep{o+b82,w+b86,vwb00}. Core emission from radio-loud QSOs tends to have
a flatter spectral index ($\alpha>-0.5$; $S_\nu\propto
\nu^\alpha$) than lobe emission due to a superposition of many
Doppler-boosted synchrotron self-absorbed spectra. Hence observing a
radio loud QSO down the jet axis, not only enhances the observed core
emission, but also flattens the radio spectrum \citep{w+b86,j+m02,j+m06}.

\begin{figure}
\centerline{\psfig{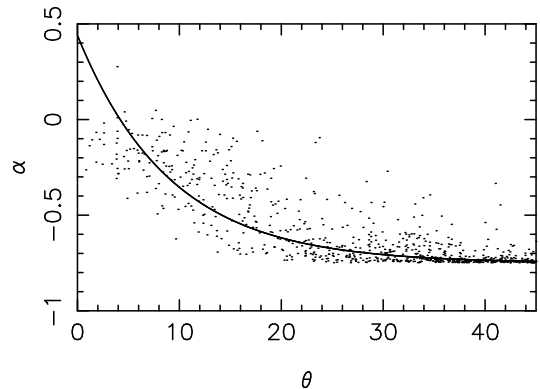}}
\caption{The 330\,MHz to 1.4\,GHz radio spectral index as a function
  of source orientation from the empirical simulation of
  Wilman~et. al.~(2008).}
\label{fig:si_ang}
\end{figure}

The relation between $\alpha$ and observing angle is by no means exact
and must include significant scatter due to intrinsic differences
between radio sources.
A guide for connecting $\alpha$ to the observing angle for a
radio-loud QSO can be found in the simulations of \citet{wilm08} based
on the dual-population models of \citet{j+w99}. In
their semi-empirical simulation of the radio source population
\citet{wilm08} included
a relativistic beaming model for sources with a random distribution of
viewing angles on the sky. Full details can be found in \citet{wilm08},
and in Fig.~\ref{fig:si_ang} we plot the 330\,MHz to 1.4\,GHz radio
spectral index against the viewing angle to the source for all FR I
and FR II objects in the simulation. Note that in these models the
extended emission is assumed to have $\alpha=0.75$ and the resulting
distribution of spectral indexes is limited by the same value. The
solid line in Fig.~\ref{fig:si_ang} is a linear fit to the data in
$\log(\alpha+0.75)$--$\theta$ space. We only plot the figure for
$\theta<45^\circ$ as objects observed from a larger angle would likely
be obscured. Steeper-spectrum objects do exist and help illustrate the
difficulties in modelling the radio source population. However, the
radio spectral index has been shown to work as an orientation
indicator \citep{w+b86,bro96}, though we must be mindful that there
is a large scatter in the relation.

By taking $R$ or $\alpha$ as proxies for orientation, and comparing these
with broad line profiles from optical spectra, several authors have found
strong evidence that broad emission lines are narrower in radio-loud
QSOs viewed down the jet axis \citep{w+b86,bro96,vwb00,j+m06}. The simple
interpretation of these results in that the BLR is confined to a
disk, hence reducing the line-of-sight velocity when viewed down the
pole. It is also worth pointing out that if the BLR was in any way
coupled to the jet, i.e. the material was entrained by the jet, then
we would expect a large offset and broadening of the emission  lines
which we do not observe. 


For radio-quiet QSOs, a common technique for studying the effect of
orientation on the BLR is by making orientation-independent
empirical estimates for the mass of QSO SMBHs. By comparing these
with virial SMBH mass estimates, which would be effected by a
non-spherical BLR, $f$ can be calculated and used to deduce
the emitting structure.

\citet{m+d02} compared SMBH mass estimates from the \Ms\ relation and
the \hbeta\ virial relation in a small sample of radio-quiet QSOs.
Based on a fit to their line width distribution, where they assume a pure-disk
model for the BLR (i.e. $v_i=0$), \citet{m+d02} calculated the average 
correction factor $f$ for virial SMBH mass estimates as a function of observed
line width. They then showed that correcting virial masses by this
factor improved the agreement with SMBH mass estimates based on the
\Ms\ relation. Hence they argue that their disk model for the \hbeta\
BLR is valid.

Since, \citet{lab06} and \citet{dec08} have produced similar analyses
indicating that the \civ\ BLR may also be disk-like.

%
%
%
%
%

Studies of the broad line distribution on its own have also shed light
on the geometry of the BLR. \citet{ost77} argued that the lack of
narrow-line Seyfert~1s in a sample of 36 objects was evidence that
rotation could not be invoked to explain the width of broad
lines. Later \citet{me2} and \citet{me3} investigated the \mgii\
and \civ\ line width distribution in a large dataset of ($\sim100,000$) QSO
spectra. These analyses showed that the variation in line width
between QSOs is so slight that the BLR cannot be strongly
geometrically constrained.

\citet{rich02} showed that \civ\ emission-line shifts with respect to
other spectral features are due to/accompanied by a drop off in flux
in the red wing of the line. They proposed an orientation-based model
to describe their observations in which the \civ\ BLR is associated
with an outflow, the far side of which can be obscured depending on
the angle to the source.

A somewhat stable empirical picture of the BLR is emerging from these
observational studies. Each of the most often studied broad lines
(\hbeta, \mgii\ and \civ) shows indications that the basic geometry
may be disk-like. However, the \mgii\ and \civ\ emission lines at
least (and likely \hbeta) cannot be confined completely to a disk and
must have some other component to their velocity field.


\subsection{Orientation and the NLR}

The NLR can be observed in type~2 AGN, and many studies into the
effect of orientation on NLR properties exploit this by comparing
type~1 and type~2 objects. The implication being that
type~1 AGN are all viewed roughly pole-on, and type~2 AGN are viewed
roughly edge-on. Therefore any differences between the NLR properties
in suitably matched samples of type~1s and type~2s can be attributed
to orientation. We note here that this paper is concerned primarily with high
luminosity QSOs and the role of luminosity on the torus in AGN and the
critical angle between type~1 and type~2 is not fully known (although
see \citealt{wil00,sim05}). Hence
results on low-luminosity Seyferts may not directly translate to quasars.

Studies comparing the flux ratio between high and low-excitation lines
indicate that type~1 AGN have a higher-ionisation NLR than type~2s
\citep{s+o81,coh83,sch98,m+t98clr}. A number of
interpretations for this result have been suggested. The most popular
calls for a stratified NLR in which higher-excitation lines are
located in a smaller region and can be
hidden by the obscuring torus when viewed edge on
\citep{m+t98clr,m+t98dc}. \citet{sch98} also argued for a second
interpretation in which type~1 AGN
have smaller NLRs, and hence are ionised by a stronger continuum.


Subsequent studies, either comparing the differences between type~1
and~2 spectra \citep{nag01,zha08}, or spatially resolving the NLR
\citep{ben06s2,ben06s1,kra09} tend to support a stratified NLR in
which higher-excitation lines are obscured when viewed edge
on. However, little work has been done in assessing the impact of
orientation in large samples of type~1 quasars. 

\begin{figure*}
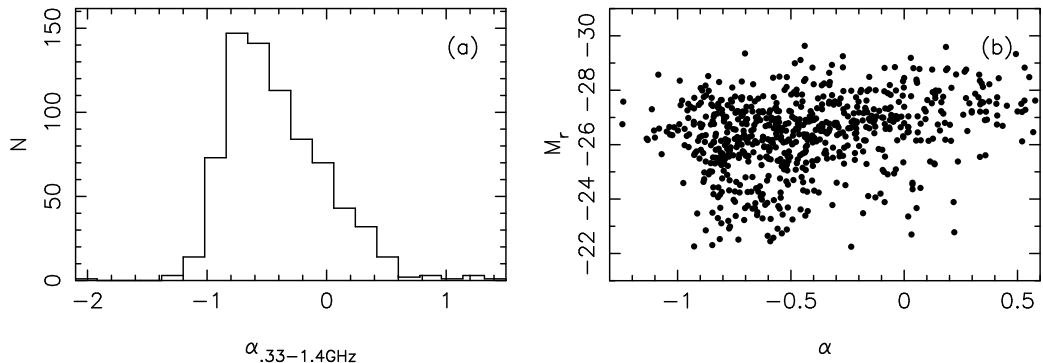

\begin{centering}
\centerline{\psfig{file=plot_si_hist.ps,width=7cm,angle=-90}
\psfig{file=plot_Mr.ps,width=7cm,angle=-90}}
\end{centering}
\caption{\empha\ The spectral index distribution of our sample as calculated
  from the 330\,MHz and 1.4\,GHz fluxes. \emphb\ The relation between
  radio spectral index and $r$-band absolute magnitude.}
\label{fig:si_hist}
\end{figure*}

\subsection{This paper}

In this paper we follow on from work presented in \citet{j+m06}
and use the radio spectral index as a proxy for orientation to
investigate the effects
of orientation on both the broad- and narrow-line region in radio-loud
quasars. In Section~\ref{sec:sample} we present the sample used in this
analysis, Section~\ref{sec:anal} describes the emission line analysis that
is new to this work and in Section~\ref{sec:results} we present our
results and discuss them in terms of models for the broad and
narrow-line regions. Throughout, this
paper we assume a flat $(\Omega_{\rm m},\Omega_{\Lambda})=(0.3,0.7)$,
$H_{0}=70\,{\rm km\,s}^{-1}\,{\rm Mpc}^{-1}$ cosmology.

\section{The Sample}
\label{sec:sample}

Both the photometric and spectroscopic optical data for this paper
come from the SDSS \citep{yor00} data release 5 (DR5; \citealt{ald07})
quasar catalogue compiled by \citet{sch07}.
Details of the SDSS telescope and spectrograph are given in
\citet{gun06} and \citet{sto02}.

In creating the SDSS QSO sample used in this paper, \citet{sch07}
visually inspect all of the candidate spectra to determine their
classification. Hence the final sample includes
all broad-line SDSS spectra, combining radio and optically selected
objects in amongst other broad-line objects selected under different
criteria. To ensure we are not biased, we include in our sample only
objects which were colour-selected as low or high-redshift QSO candidates
\citep{rich02b} as the BEST target. And to ensure accurate radio
identification we take only those objects which lie in the Faint
Images of the Radio Sky (FIRST; \citealt{bwh95}) region, and have a
FIRST detection (6255 objects).



This core sample was then cross matched with the WENSS catalogue
\citep{ren97} to produce 746 matches. These were then matched to the
NVSS \citep{con98} for which we found all sources had a unique match.

The result is a sample of broad-line galaxies from the SDSS
with WENSS and NVSS fluxes for each. Note that, while 
all of the WENSS detections have an NVSS detection
there are many NVSS sources in the cross-matching area with no WENSS
detection, due to the relative depths of the surveys. Furthermore the
requirement of a broad line in the spectrum means that contamination
from BL Lac objects, that may be strongly beamed in the optical, is
low. In fact taking the classic broad-line equivalent width cut of
5\,\AA\ (e.g. \citealt{u+p94}) to define BL Lacs there are none in our sample.

Fig.~\ref{fig:si_hist}\empha\ shows the 330\,MHz to 1.4\,GHz radio
spectral index distribution of our
sample. In this paper we are taking radio spectral index as an
orientation indicator. The objects with flatter spectral indices
are considered to be viewed more pole-on than
the sources with steeper radio spectra. Fig.~\ref{fig:si_hist}\emphb\
shows how the $r$-band \citep{fuk96} absolute magnitude of our sample
depends on radio spectral index. The correlation (a Spearman
rank test gives $r_s\sim-0.23$ with
$P(r_s)\ll1$\,\%) evident between
$\alpha$ and $M_r$ may indicate a degree of obscuration of the
accretion disk in QSOs viewed from an angle away from the pole
(however see Fig.~\ref{fig:osi_si} and discussion later in
paper). Furthermore the correlation in Fig.~\ref{fig:si_hist}\emphb\
may be indicative of a selection bias in our sample. Since we find
brighter objects have steeper spectral indexes the optical selection of
the SDSS sample may bias us towards these (potentially more pole-on)
sources. In the analysis that follows we will try to correct for
potential luminosity bias.

\section{Emission line analysis}
\label{sec:anal}

In this paper we will compare a series of emission line parameters
with radio spectral index to infer conclusions about the emitting
regions. \citet{me2} and \citet{me3} have already performed an
analysis of all of the \mgii\ and \civ\ line in our sample and we will
use the parameters for these lines derived in those papers. In
addition we will be measuring parameters of the narrow \oii\
$\lambda3727$ and \oiii\ $\lambda5007$ lines.

\begin{figure}
\centerline{\psfig{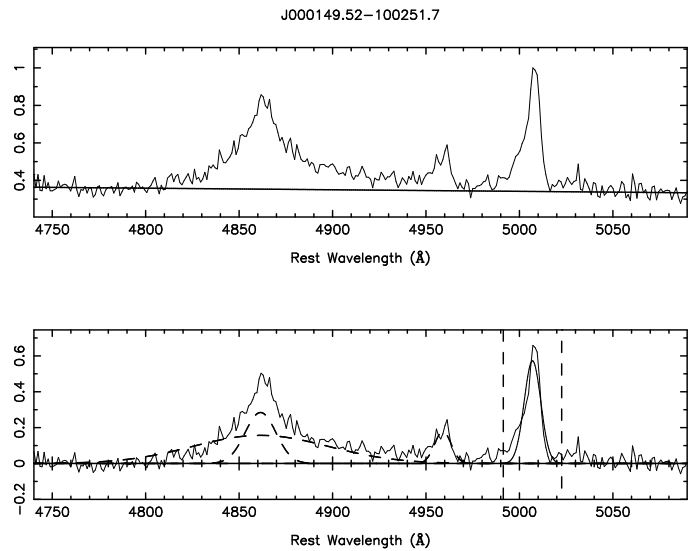}}
\caption{Example of a fit for the \oiii\ line for QSO
  J000149.52-100251.7. The top plot shows the original spectrum and
  linear continuum fit. The second plot shows the continuum subtracted
  spectrum along with Gaussians fitted to the \hbeta\ and \oiii\
  $\lambda4960$  lines (dashed), and the \oiii\ $\lambda5007$ line
  (solid). Vertical lines indicate the $\pm1.5$FWHM limits within
  which we sum the flux. The $y$-axis in each plot gives the
  flux density normalised to the peak in the top plot.}
\label{fig:plot_eg}
\end{figure}

\oii\ is not blended with any other emission lines and is therefore
relatively straightforward to analyse. We
fit a quadratic continuum between windows either side of the
line and subtract it from the spectrum.  The resolution of the spectra
does not require us to fit the \oii\ doublet as two individual emission
lines, therefore a single Gaussian is fitted to the 
line to define its extent. We
then sum the flux in the continuum-subtracted spectrum within
$\pm1.5\times$ the FWHM of the fitted Gaussian. 

\oiii\ is a doublet, and can be blended with the wing of the broad
\hbeta\ line. We fit a continuum between the extremities of the
\hbeta\ line and subtract it from the spectrum. We then fit a
double-Gaussian model to \hbeta\ and 
single Gaussians to the \oiii\ $\lambda4960$ and $5007$ doublet
fixing their relative amplitude to 1:3. The three Gaussians not
pertaining to \oiii\ $\lambda5007$ are subtracted from the data and the
flux was summed between $\pm1.5\times$ the FWHM of the Gaussian fit
to the line.

Note that this method for fitting the \oiii\ lines does not always
give accurate fits to the whole \hbeta\ region of the
spectrum. Differing \hbeta\ profiles, continuum shapes and varying
iron emission in the region all have an effect. However, out fitting
method does accurately correct for contaminants to the \oiii\ $\lambda5007$
line. However, to make sure the results were acceptable, all fits were
inspected manually and in some cases the continuum fit had to be
adjusted to obtain adequate fits. An example fit is shown in
Fig.~\ref{fig:plot_eg}.

\section{Results}
\label{sec:results}

Here we compare various emission line
parameters with the radio properties of our sample in order to derive
conclusions about the line emitting regions in QSOs. Since the BLR and
NLR differ in size by many orders of magnitude we will consider them
separately.

\begin{figure*}
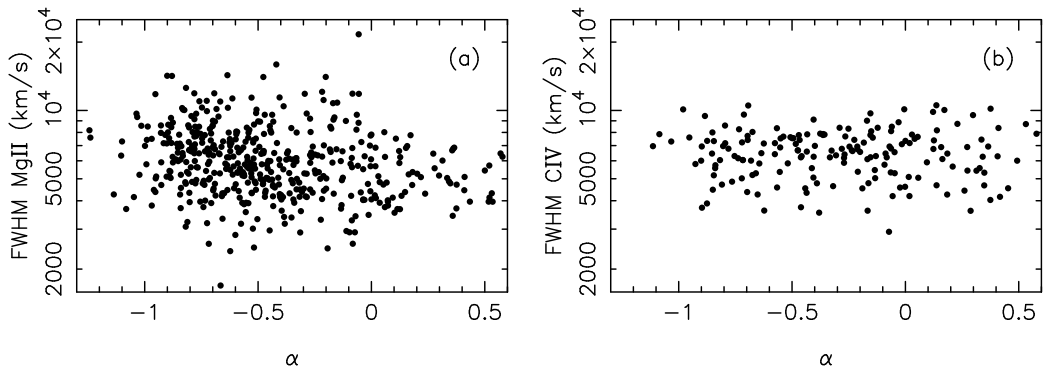

\begin{centering}
\centerline{\psfig{file=plot_mgii_vw.ps,width=7cm,angle=-90}
\psfig{file=plot_civ_vw.ps,width=7cm,angle=-90}}
\end{centering}
\caption{These figures show \mgii\ \empha\ and \civ\ \emphb\ line
  widths as a function of radio spectral index, $\alpha$. While there
  is considerable scatter in \empha\ there is a trend such that
  flatter spectrum sources have narrower line widths. \emphb\ shows no
  evidence for a trend (see text for details).}
\label{fig:lw_si_ind}
\end{figure*}

\subsection{The BLR}

We are interested in how broad lines in quasar spectra are effected by
orientation. In particular, previous studies have shown evidence that
emission lines are broader in objects observed edge on, indicating
a disk-like BLR. We are using the radio (330\,MHz to 1.4\,GHz)
spectral index as a proxy for orientation in our sample, and in
Fig.~\ref{fig:lw_si_ind} we show \mgii\ \empha\ and \civ\ \emphb\
emission line widths plotted against the radio spectral index.

In each panel of Fig.~\ref{fig:lw_si_ind} there is a significant
amount of scatter. However, it is apparent that
there is an anticorrelation between \mgii\ line width and $\alpha$ in
Fig.~\ref{fig:lw_si_ind}\empha, most evident
for $\alpha<-0.2$. A Spearman rank test on the data in
Fig.~\ref{fig:lw_si_ind}\empha\ gives $r_s\sim-0.26$
($P(|r_s|>0.26)\ll 0.01$\,\%
given a random distribution). On the other hand
Fig.~\ref{fig:lw_si_ind}\emphb\ shows no evidence for a correlation
($r_s\sim-0.02$; $P(r_s)\sim0.79$). There are fewer objects with a
\civ\ line width measurement and we test whether this is reducing the
significance of any correlation with respect to \mgii. We resample the
\mgii\ line width distribution randomly to have the same number of
objects as have \civ\ measurements and perform Spearman rank tests in
each of the subsamples. In all cases we found
$P(r_s)<0.05$ and $P(r_s)<0.01$ in 80\,\% of
the subsamples.

\begin{figure*}
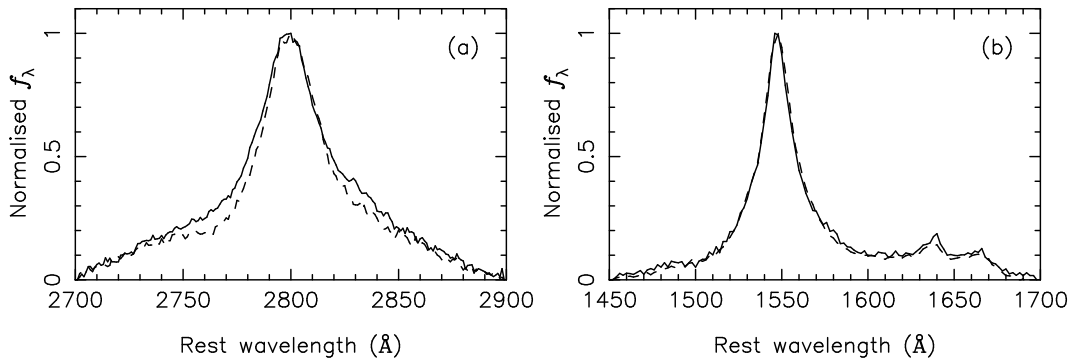

\begin{centering}
\centerline{\psfig{file=mg_comp3.ps,width=7cm,angle=-90}
\psfig{file=civ_comp3.ps,width=7cm,angle=-90}}
\end{centering}
\caption{Composite spectra of the \mgii\ \empha\ and \civ\ \emphb\
  line regions for objects with $\alpha<-0.5$ (solid) and
  $\alpha>-0.5$ (dashed). In \empha\ there is a clear difference
  between the two lines with the flat spectrum sources giving a
  narrower emission line. In \emphb\ the \civ\ line shows very little
  difference between the flat and steep spectrum sources. } 
\label{fig:lw_si_comp}
\end{figure*}

To highlight how the profile of the lines depends on
$\alpha$, Fig.~\ref{fig:lw_si_comp} shows median-value
(variance-weighted composites show the same results) composite
spectra of the \mgii\ \empha\ and \civ\ \emphb\ line regions for
objects that have $\alpha<-0.5$ (solid) and $\alpha>-0.5$ (dashed). In
each figure the spectra have had a continuum subtracted between the
red and blue ends of the plot, and have had their flux densities
normalised to their maximum values for comparison.

The correlation in Fig.~\ref{fig:lw_si_ind}\empha\ reproduces the
results of \citep{j+m06} and is in agreement with other studies
indicating a flattened BLR for the \mgii\ line.
The lack of any correlation in Fig.~\ref{fig:lw_si_ind}\emphb\ may be
of more consequence. The indication is that the \civ\ BLR is, to a
large extent, isotropic and unaffected by orientation effects. This is
in contrast to previous studies (e.g. \citealt{vwb00,dec08}).

\subsubsection{Discussion}

Our results for the \mgii\ line are in line with previous results. The
implication is that the \mgii\ emission region has a significant
velocity component in the plane of the disk, potentially orbital
motions. However, while the difference between the composite spectra in
Fig.~\ref{fig:lw_si_comp} is highly significant, the magnitude of the
difference is not huge. This indicates that a thin-disk
model for the BLR is not sufficient and some other, non-planar,
component to the motion is required.

We find no evidence for orientation effects in the \civ\
line. Previous authors have reported an effect
(e.g. \citealt{vwb00,dec08}) and we discuss the potential reasons for
the discrepancy here.

\citet{vwb00} compared the radio core-to-lobe
flux ratio $R$ with \civ\ line widths for 37 radio-loud QSOs. They
found no correlation between the FWHM of the \civ\ line and
$R$. However, they found that line width measurements that were more
weighted towards the wings of the line (the full width at 20\,\%
maximum, and inter-percentile velocity, IPV, widths) did show a significant
anti-correlation with $R$. Given the small sample size in
\citet{vwb00}, and the difficulties they had in finding a correlation,
it may be that their result is not robust.



\citet{dec08} follow a method laid out by \citet{m+d02} in which
virial SMBH mass estimates are compared to empirical mass estimates
from the Magorrian relation \citep{mag98}. The argument follows that
residuals between the two mass estimates can be ascribed to
orientation effects. Assuming a disk model for the BLR \citet{m+d02}
derive the average correction factor for virial mass estimates ($f$ from
equation~\ref{equ:vir_f}; note that \citealt{m+d02} define $f$ as the
square root of $f$ as we define it in equation~\ref{equ:vir_f}) as a
function of observed FWHM. Put simply, in their model an emission line
that is observed to be narrow is more likely to be observed pole-on,
and hence effected strongly by orientation,
than an emission line that is observed to be broad. Hence narrower
lines will, in general, have a higher value of
$\sqrt{f}$~($=1/(2\sin\theta)$ in their model).

Both \citet{m+d02} and \citet{dec08} find that this disk model for the
BLR give good fits to the observed residuals between virial SMBH mass
estimates and those from galaxy scaling relations. Furthermore, the
disk model gives an improved fit 
compared to an isotropic model in
which $f=const$. However, we argue here that this is not necessarily
evidence for a purely disk-like BLR.

\begin{figure}
\begin{centering}
\centerline{\psfig{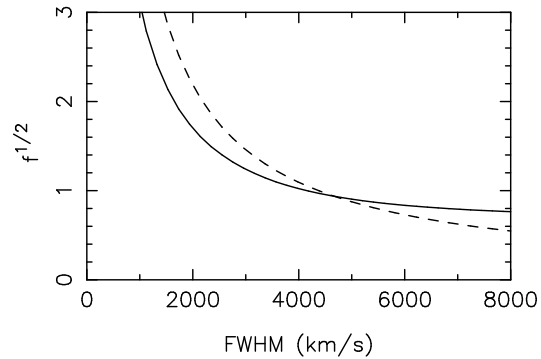}}
\end{centering}
\caption{Here we compare the disk model from McLure \& Dunlop~(2002),
  that connects the observed FWHM of a line to the average correction
  factor $f$ for a virial SMBH mass estimate (solid line; note McLure \&
  Dunlop~(2002) define $f$ as the square root of the parameter in
  equation~\ref{equ:vir_f}, we plot $\sqrt{f}$ to present a simpler
  comparison to their data), to a model in which the FWHM of lines is
  unrelated to orientation or SMBH mass (dashed line). The two models
  show almost identical behaviour indicating that data which follow the
  trend of these lines is not necessarily evidence for a disk-like
  BLR.}
\label{fig:m+d_rej}
\end{figure}

In recent years some concern has grown up over the accuracy of virial
SMBH mass estimators. In particular \citet{me3} argued that there is
not enough dynamic range in either \mgii\ or \civ\ line widths for
them to be useful in virial estimators given the typical uncertainty
of the mass estimates. If we assume that emission line width does not
help in mass estimation, and is essentially random, then narrower
lines will bias mass estimates low, and broad lines will bias mass
estimates high. The result, in terms of derived values for $f$ as a
function of observed FWHM, is extremely similar to the disk model of
\citet{m+d02}. In Fig.~\ref{fig:m+d_rej} we show the best fit model from 
\citet{m+d02} which assumes $\overline{V}_{orb}=4375$\,km/s,
$\sigma_{orb}=1400$\,km/s and $\theta_{max}=47^\circ$ as the
solid line. We have then taken a model in which observed line widths
are unrelated to the virial mass. That is, we have taken the same
$\overline{V}_{orb}$ as \citet{m+d02} and assumed that the FWHM is
independent of the virial velocity of the BLR, hence,
$\overline{f}(FWHM)=\overline{V}_{orb}/FWHM$. This relation is plotted 
in Fig.~\ref{fig:m+d_rej} as the dashed
line. Note that our line is not a fit to data, we simply take
parameters from the \citet{m+d02} fit which assumes a different
model. However, the two lines show the same trend. Hence, we argue
that the residuals 
\citet{dec08} find between their SMBH mass estimates may be due more 
to a failing in the virial mass estimators than a disk-like BLR.

There may be some concern that the \mgii\ and \civ\ lines occur in
spectra of QSOs in differing redshift ranges, and hence different
luminosity ranges. However, studies of the widths of these emission
lines in large samples of QSOs have shown little-to-no dependence on
luminosity or redshift \citep{corb03,me2,me3}. As a further test we
restrict our analysis to only those (88) objects in our sample that
have both \mgii\ and \civ\ in their spectra. Correlating the line
widths of these with their radio spectral index we still find a
significant correlation ($r_s\sim-0.3$; $P(r_s)<0.01$) for \mgii\ and
none for ($r_s\sim-0.1$; $P(r_s)\sim0.2$) \civ.

\subsection{The NLR}

\begin{figure}
\begin{centering}
\centerline{\psfig{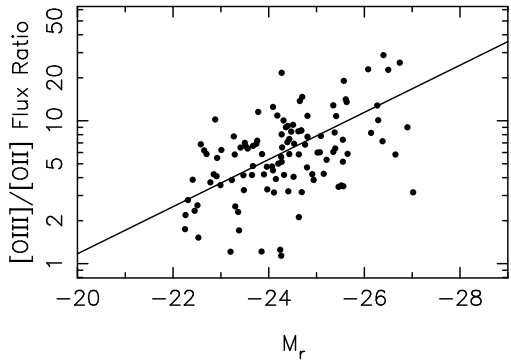}}
\end{centering}
\caption{The flux ratio between the \oiii\ and \oii\ lines plotted
  against the $r$-band absolute magnitude of the object. The clear
  correlation indicates that the NLR in brighter sources is in a
  more highly ionised state. The solid line is a minimum $\chi^2$ fit.}
\label{fig:Or_Mr}
\end{figure}

The ratio between the flux of the \oiii\ $\lambda$5007 and \oii\
$\lambda$3727 lines gives a measure of the ionisation level of the
NLR. One of the primary drivers of this is the intrinsic luminosity of
the source and in Fig.~\ref{fig:Or_Mr} we show how the ratio depends
on the $r$-band luminosity of the QSO.

\begin{figure*}
\begin{centering}
\centerline{\psfig{file=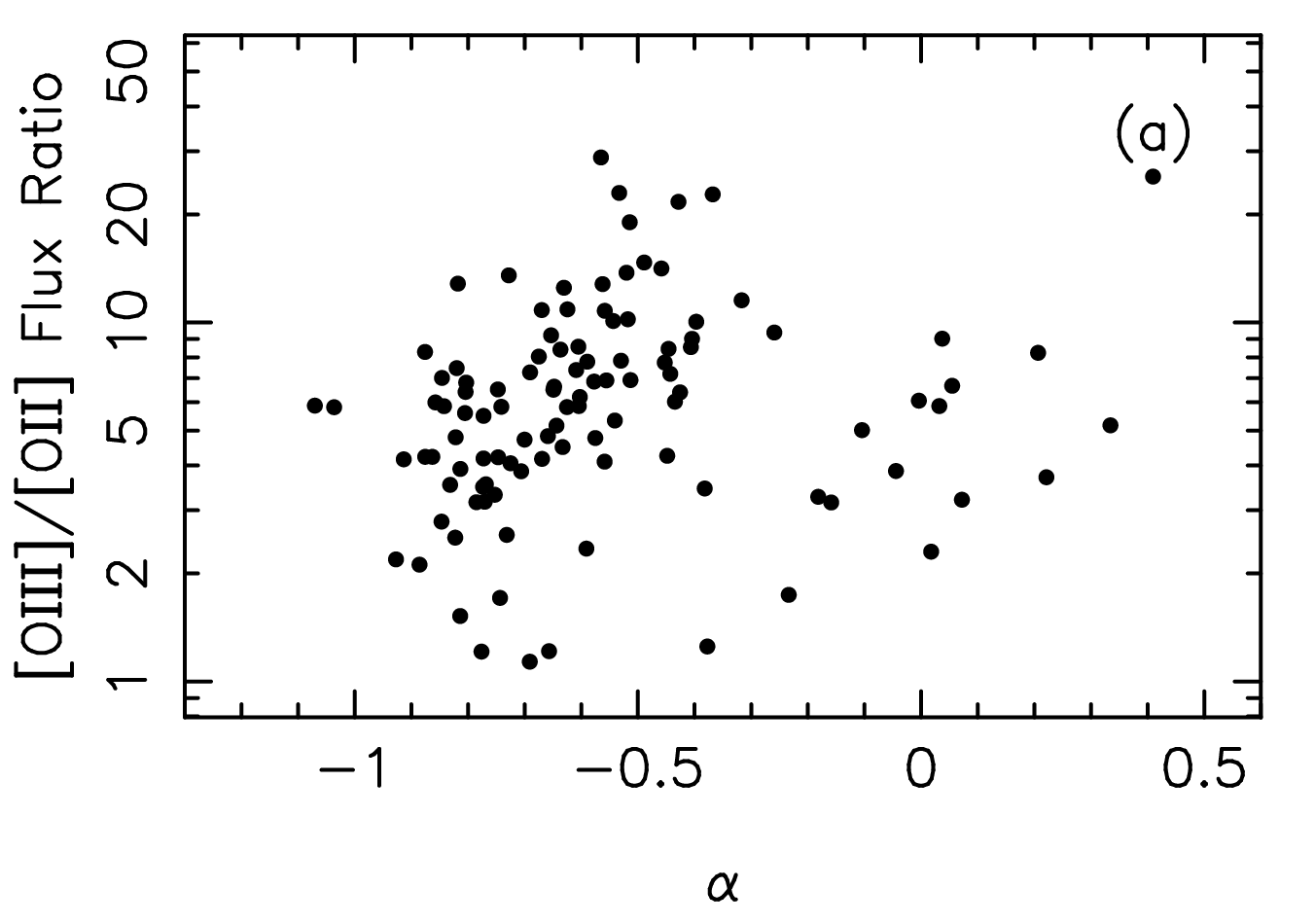,width=7cm,angle=-90}\hspace{0.5cm}\psfig{file=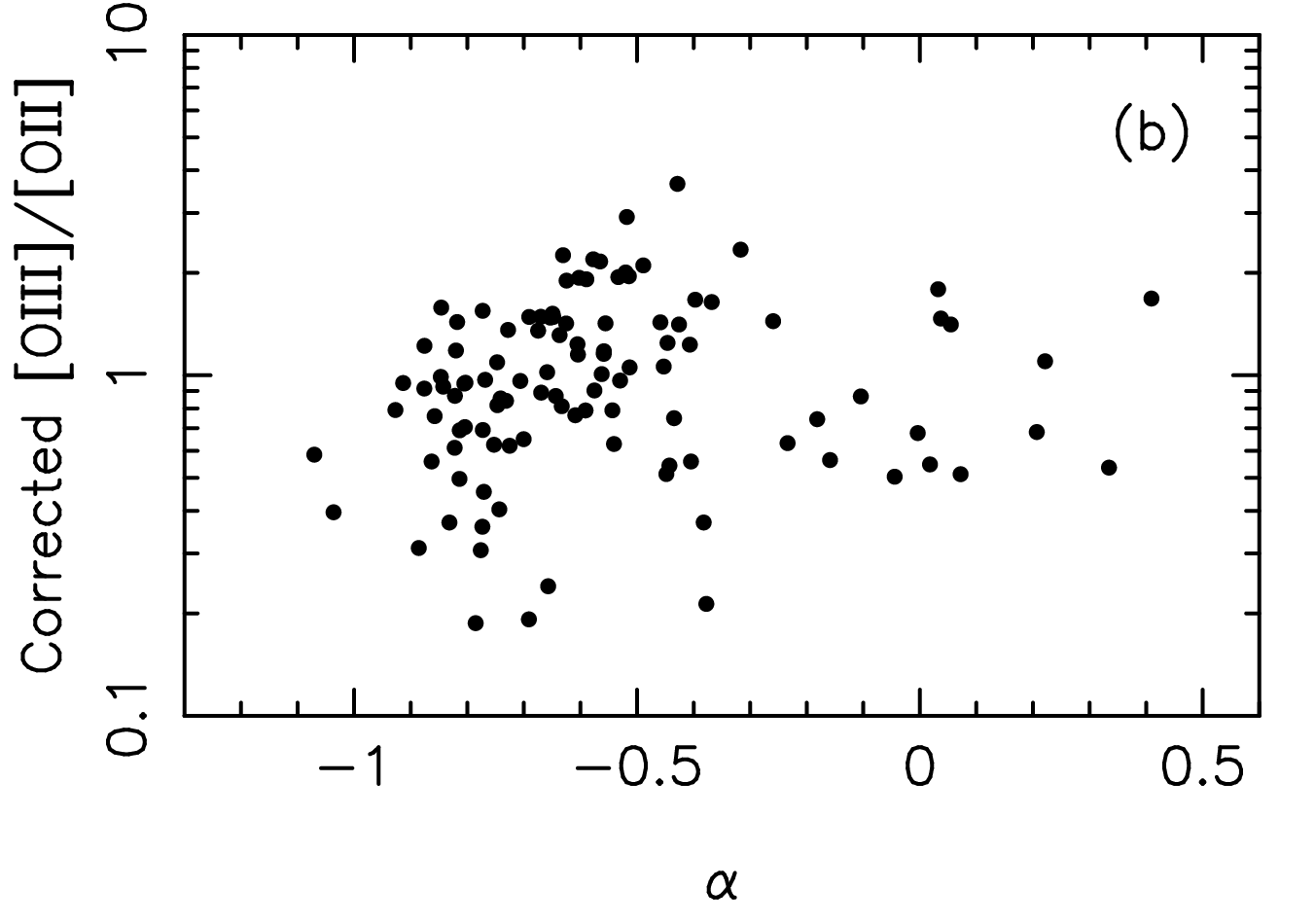,width=7cm,angle=-90}}
\end{centering}
\caption{The flux ratio between the \oiii\ and \oii\ lines plotted
  against the radio spectral index for the QSOs in our sample. There
  is a clear positive correlation indicating that the NLRs of more pole-on QSOs
  have a higher excitation level. There is an indication of a bimodal
  distribution in $\alpha$ in this plot with an increased number of
  objects with $\alpha\sim0$. This is likely due to Doppler boosting
  of sources from the highly-populated fainter end of the luminosity
  function. In \emphb\ we correct for potential luminosity bias using
  the best fit line in Fig.~\ref{fig:Or_Mr}.}
\label{fig:Or_si}
\end{figure*}

Previous studies of the NLR have found that high excitation lines are
emitted from a smaller region when compared to low excitation
lines. Furthermore, when viewed edge-on high-excitation lines are less
luminous, potentially due to absorption close to the nucleus. To
test the effect of orientation on the NLR in our sample of QSOs we
plot the \oiii\ to \oii\ flux ratio
against radio spectral index in Fig.~\ref{fig:Or_si}\empha. To check
that the results are not due to luminosity bias
we correct for the correlation evident in Fig.~\ref{fig:Or_Mr} by
dividing the measured flux ratio by the value of the best fit relation
at that magnitude. These corrected results are shown in
Fig.~\ref{fig:Or_si}\emphb.

\begin{figure*}
\begin{centering}
\centerline{\psfig{file=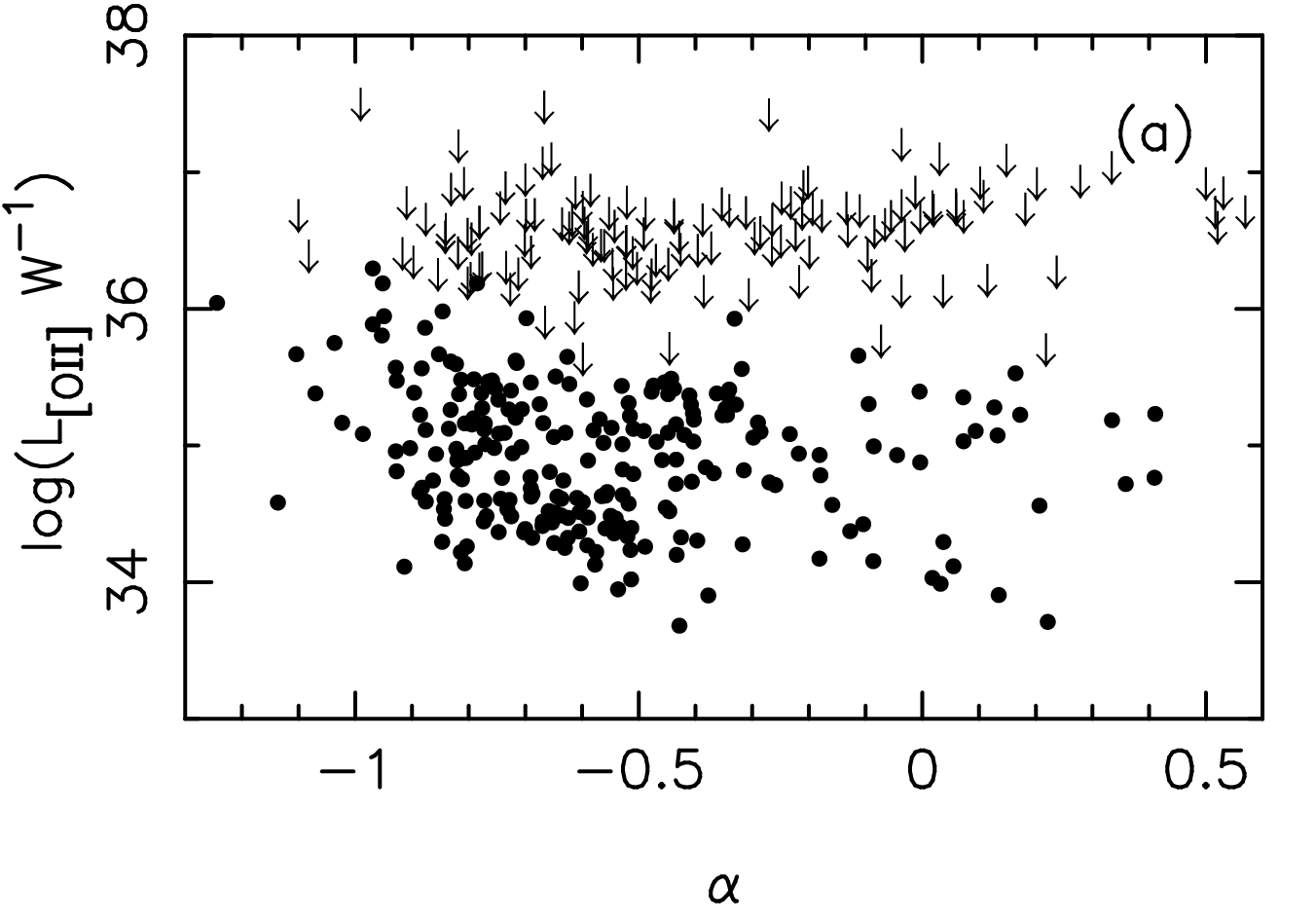,width=7cm,angle=-90}
\psfig{file=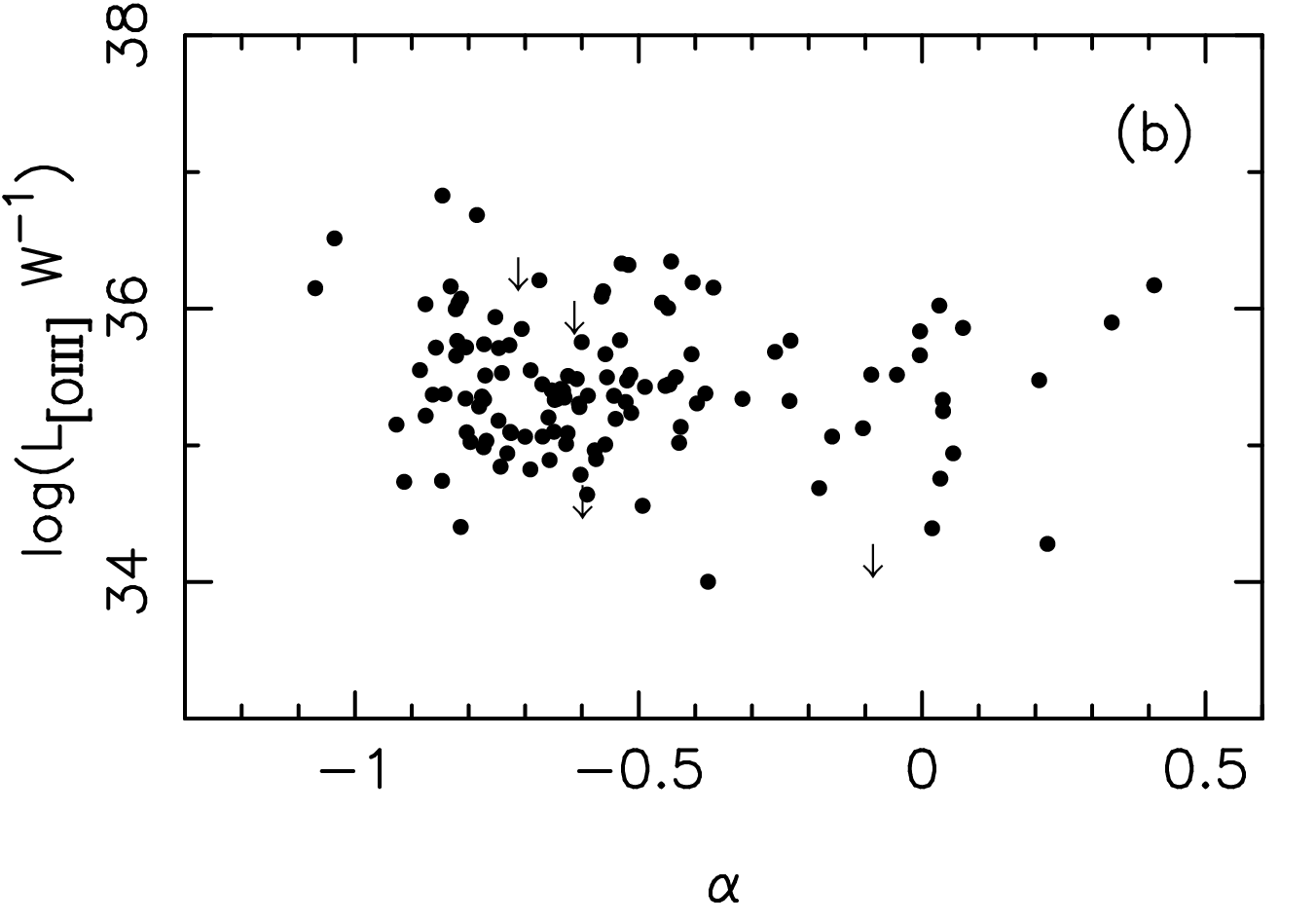,width=7cm,angle=-90}}
\end{centering}
\caption{The luminosity of the \oii\ $\lambda$3727 \empha\ and \oiii\
  $\lambda$5007 \emphb\ lines plotted against the spectral index of
  the source. For spectra in which the emission line was not visible an
  upper limit is plotted calculated from the spectral S/N. Both
  \empha\ and \emphb\ show the same behaviour with line flux
  anti-correlated with $\alpha$. This is primarily due to a lack of
  objects with flat spectral indices, i.e. large $\alpha$, and bright
  emission lines.} 
\label{fig:Ofl_si}
\end{figure*}

At face value the correlation apparent in Fig.~\ref{fig:Or_si} is
exactly as we expect. For flat spectrum pole-on QSOs there is less
absorption of the inner high-excitation NLR and the \oiii/\oii\ ratio
is increased. However, if we look at the fluxes of \oii\ and \oiii\
individually we find that the picture is more
complicated. Fig.~\ref{fig:Ofl_si} shows the \empha \oii\ and \emphb \oiii\
line luminosities plotted against the spectral index of the QSO.

In both Fig.~\ref{fig:Ofl_si} \empha\ and \emphb\ the same behaviour
is observed. There is an anticorrelation between the line flux and
spectral index implying that, on average, more pole-on QSOs
have fainter oxygen lines. Taking $\alpha=-0.5$ as the approximate
mid-point of the spectral index distribution, those objects with
steep-spectra ($\alpha<-0.5$) 
have $\sim$1.3 times the \oiii\ flux and $\sim$1.4 times
the \oii\ flux compared with the flat-spectrum  ($\alpha>-0.5$)
objects. This is not easy to 
square with our interpretation of Fig.~\ref{fig:Or_si} above: that the
reduced oxygen flux ratio for steep spectrum sources is due to absorption.

\begin{figure}
\begin{centering}
\centerline{\psfig{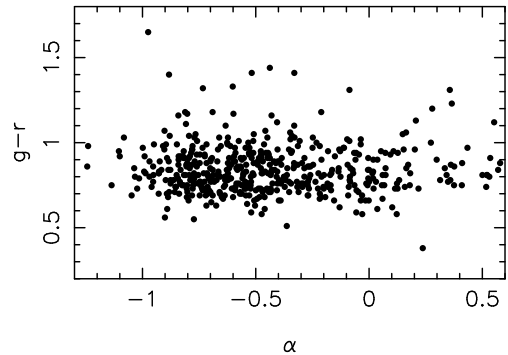}}
\end{centering}
\caption{$g-r$ colour of our sample correlated with radio spectral
  index. There is no evidence for a correlation here, indicating that
  QSOs viewed from an angle are not heavily reddened compared to those
  viewed pole-on.}
\label{fig:osi_si}
\end{figure}

One possibility is that the flat-spectrum QSOs are intrinsically
fainter in the optical, and therefore produce fewer ionising
photons. While Fig.~\ref{fig:si_hist} shows that the
observed optical magnitudes are, in fact, brighter for the flat
spectrum QSOs. It may be that in the steep-spectrum QSOs, that are
observed from a larger angle, may suffer from more extinction in the
optical. Extinction should also redden the optical spectrum of our
sample, and in Fig~\ref{fig:osi_si} we show the $g-r$ colour of our
sample as a function of radio spectral index.



We find no evidence for a correlation in Fig.~\ref{fig:osi_si}
(a Spearman rank test gives $r_s\sim-0.03$ with
$P(r_s)\sim0.4$). Since the optical bands sample different rest
wavelengths at different redshift we check that differing redshift
distributions for low and high $\alpha$ were not responsible for the
lack of correlation by also measuring the optical spectral index off
the SDSS spectra. This is an inherent part of our line fitting
procedure and we find that, regardless of which line we measure the
optical spectral index under, we find no correlation with radio
spectral index. It is also worth noting that we also find no evidence
for increased reddening in the flat-spectrum population, which could
arise due to synchrotron emission associated with the jet. 

The implication is that the QSOs in our sample that are viewed from
an angle are 
not heavily reddened compared to those viewed pole-on. Hence we find it
difficult to explain Figs.~\ref{fig:Or_si} and~\ref{fig:Ofl_si} through
dust extinction.

\begin{figure}
\begin{centering}
\centerline{\psfig{file=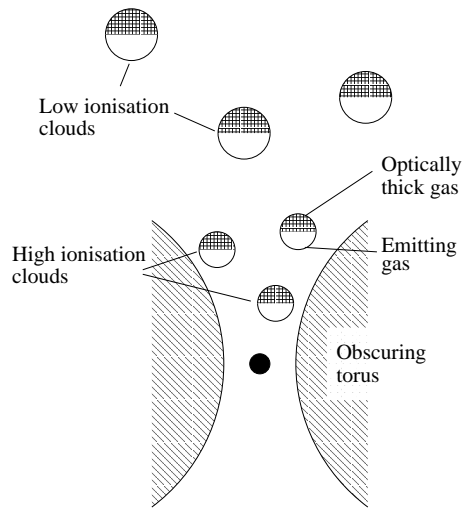,width=6cm}}
\end{centering}
\caption{A schematic showing the NLR model discussed in the text. The
  central source ionises the face of optically thick clouds in the
  surrounding galaxy. In the inner NLR the radiation field is stronger
  and higher ionisation lines are brighter compared to the outer
  NLR. When observed from an angle a large portion of the ionised
  surface of each cloud is visible and the lines are bright, however,
  the inner high-excitation NLR is obscured by the torus. When
  observed from the pole the inner NLR is not hidden by the torus, but
  we are looking through the back of the NLR clouds and see less line
  emission in total.}
\label{fig:nlr_model}
\end{figure}

Another potential explanation for Figs.~\ref{fig:Or_si}
and~\ref{fig:Ofl_si} is with a model in which the narrow-line emitting
clouds of the NLR are themselves optically thick. In this model only the
surface of the clouds facing the central ionising source would be
ionised and, when observed from behind, the ionised region would not be
visible through the cloud. Fig.~\ref{fig:nlr_model} shows a schematic
of this model.

In the model presented in Fig.~\ref{fig:nlr_model} the NLR exists on
the face of optically thick clouds ionised by the central
source. In the inner NLR, the radiation field is stronger,
and higher ionisation lines are brighter compared to the outer
NLR. When observed from an angle a large portion of the ionised
surface of each cloud is visible and the lines are bright, however,
the inner high-excitation NLR is obscured by the torus. When
observed from the pole the inner NLR is not hidden by the torus, but
we are looking through the back of the NLR clouds and see less line
emission in total. This model can simultaneously explain the relation
between the \oiii/\oii\ ratio and radio spectral index, and the trend
between narrow-line flux and $\alpha$.

\subsubsection{Discussion}

In the previous section we have shown that the optical narrow-line
strengths of QSOs correlate with their radio spectral index that we
employ as an orientation indicator.
Narrow line strengths are commonly employed to
estimate bolometric luminosities of AGN \citep{heck04,sim05}, as well
as being extensively used in diagnostic analyses such as BPT diagrams
\citep{bpt81}. Here we discuss the implications of our findings with
respect to these analyses.

To quantify the potential effects of orientation we split our sample
by the rough midpoint of the radio spectral index distribution
($\alpha=-0.5$). We find that the differences in average line luminosity
between the low and high $\alpha$ bins are $\sim0.17$\,dex and
$\sim0.12$\,dex for \oii\ and \oiii\ respectively. These values give
the approximate lower bound for the effect of orientation on observed
line strengths.

In fact, the effect of orientation is likely to be considerably more
than quoted above since the considerable scatter in the relationship
between $\alpha$ and QSO orientation (e.g. Fig.~\ref{fig:si_ang}) has
the effect of smoothing the observed
relationship. Quantifying the bias introduced through taking $\alpha$
as a proxy for orientation is difficult due to the many
unknowns. However, we can estimate the magnitude of the effect through a
simple simulation.

We start by producing a rectangular radio spectral index distribution
with equal probability for $-1<\alpha_{ns}<0.$. We use the subscript
$_{ns}$ to indicate that for this original $\alpha$ distribution we
are assuming there is no scatter in the $\alpha$--orientation
relationship. We then produce an observed radio spectral index
($\alpha_{obs}$) distribution by simply adding random Gaussian noise
to these values at the level found in Fig.~\ref{fig:si_ang}
(rms$\sim0.25$). We also produce narrow-line luminosities
from the $\alpha_{ns}$s with a linear
relation, including Gaussian scatter, in which the gradient,
zero-point and level of scatter are varied so that the resulting
$\alpha_{obs}$--line luminosity correlation resembles those observed in
Fig.\ref{fig:Ofl_si}.

Splitting our simulated sample by $\alpha_{obs}=-0.5$ and
$\alpha_{ns}=-0.5$ and comparing the average line luminosity in the
low and high $\alpha$ bins for each split we can obtain an estimate
for the level of bias in the results quoted above. In the case of
\oii, where we found a 0.17\,dex offset between the bins separated by
$\alpha_{obs}=-0.5$, we estimate the unbiased offset to be
$\sim0.25$\,dex. In the case of \oiii, where we found a 0.12\,dex
offset between the bins separated by $\alpha_{obs}=-0.5$, we estimate
the unbiased offset to be $\sim0.20$\,dex. That is, \oii\ lines are
roughly twice as strong (and \oiii\ lines $\sim$1.6 times as strong)
in QSOs observed from an angle, when compared with those observed pole on.

\oiii\ line luminosity has been employed to estimate the bolometric
luminosity of AGN, most often in type-2 objects in which the core is
obscured \citep{heck04,lam09}. However, the bolometric corrections
must be calibrated against unobscured type-1 objects. Extrapolating
our results from QSOs to type-2s, it may be that orientation effects
result in the \oiii\ luminosities of type-2 objects being $\gtrsim1.6$
times the value of objects observed pole-on. The result would be that
bolometric luminosities estimated for these objects would be
similarly overestimated, as well as Eddington ratios
in the examples of \citet{heck04} and \citet{lam09}.

Our results also impact comparisons between \oiii\ luminosity
functions between type-1 and type-2 AGN (e.g. \citealt{sim05,rey08}),
and the conclusions drawn from these comparisons. \citet{sim05} find
the type-1 \oiii\ luminosity function is consistently lower than the
type-2 luminosity function in their sample of SDSS AGN. However, if we shift
their type-1 luminosity function to higher luminosities by a factor of
1.6 we find that, for \oiii\ luminosities $>2-3\times10^{34}$\,W the
type-1 luminosity function is higher than the type-2. This then
impacts their estimation of the type-1/type-2 AGN fraction, increasing
it significantly.

Since we find that the ratio between \oii\ and \oiii\ line
luminosities is also affected by orientation, it may be that
the accuracy of BPT diagrams is limited by orientation effects. We
find the \oii/\oiii\ flux ratio rises by at least a factor of two with
increasing $\alpha$. \oii\ has a comparable ionisation potential to
H{\,\sc ii}, and so we may expect similar variability in the
\oiii/\hbeta\ ratio: the y-axis on the standard BPT
plot. Hence more pole-on sources would be biased low in standard
BPT plots, making them appear more like star-forming galaxies.

\subsection{BALs}

The nature of broad absorption
systems in QSOs is still under debate. A common model
assumes that all QSOs have a BAL outflow, most likely
accelerated equatorially, and only in those objects
observed through the outflow do we see a BAL
(e.g. \citealt{wey91}). Spectropolametric analysis have provided evidence to
back up this picture \citep{coh95,g+m95}. However, more recent
studies using radio properties to gauge the viewing angle to the
source have complicated the issue. \citet{bec00} showed that radio
selected BAL quasars can display compact radio morphologies and
possess both steep and flat radio spectra indicating no preferential
orientation (see also \citealt{zho06,doi09}).

There are relatively few \mgii\ BALs, while \civ\ exhibits BALs in
$\sim20$\,\% of objects \citep{tru06}. We therefore look back at all
the \civ\ lines in our sample and categorise them as BAL or non-BAL
objects by eye. Of 240 QSOs with the \civ\ line in the spectrum we
find 25 objects which display definite BALs and 171 which definitely
do not. In the remaining 44 objects there is some indication of
absorption but the S/N is such that we cannot be sure of a BAL.

\begin{figure}
\begin{centering}
\centerline{\psfig{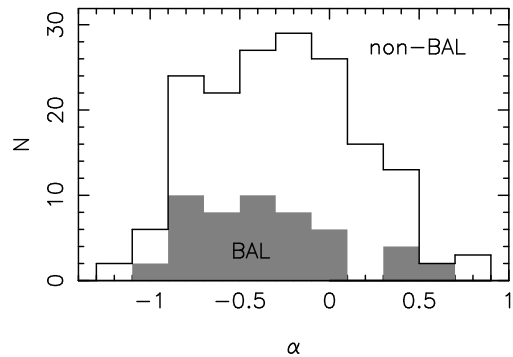}}
\end{centering}
\caption{The radio spectral index distribution of \civ\ BAL QSOs and
  \civ\ non-BALs. The height of the shaded histogram has been
  multiplied by two to highlight the shape of the BAL distribution. We
  find little difference between the two distributions}
\label{fig:bal_si}
\end{figure}

Fig.~\ref{fig:bal_si} shows the spectral index distributions of the
definite BAL and non-BAL samples, to highlight the shape of the BAL
distribution the height of the shaded histogram has been multiplied by
two. We find that these distributions are not significantly different
with respect to a KS test ($D_{ks}=0.17$; $P(D_{ks})=0.5$), and our
results are not altered if we include the objects which have
indefinite BAL identifications. However, our numbers are small here
and we cannot definitely say whether the orientation
orientation distribution of BAL vs non-BAL objects is the same.

\section{Conclusions}
\label{sec:conc}

We use the SDSS, NVSS and WENSS to define a sample of radio-loud QSOs
and measure their 330\,MHz to 1.4\,GHz spectral indexes. Using the
radio spectral index as a proxy for orientation towards the QSOs in
our sample we investigate the effect of orientation on their optical
spectra.

In our sample the \mgii\ line is significantly broader for objects with
steep spectral indexes indicating disk-like velocities for the \mgii\
BLR. However, we find no such correlation for the \civ\ line
potentially indicating a more isotropic high-ionisation BLR.

Both the \oii\ and \oiii\ narrow lines have more
flux in steep spectrum sources while the \oiii/\oii\ flux ratio is
lower in these sources. To describe both of these effects we propose a
simple geometric model in which the NLR exists primarily on the surface
of optically thick clouds facing the active nucleus and the NLR is
stratified such that higher-ionisation lines are found preferentially
closer to the nucleus. We discuss the implications of our finding on
analyses that employ narrow line fluxes (e.g. estimating bolometric
luminosities/BPT diagrams) and show that orientation can skew results
by as much as a factor of two.

We find that QSOs with flatter spectral indexes tend to be
brighter. However, we find no indication of reddening in
steep-spectrum QSOs to indicate dust obscuration of the accretion
disk as a possible explanation, or in the flat-spectrum population
which would imply contamination by synchrotron emission from the
jet. Finally we find no evidence that BAL 
QSOs have a different spectral index distribution to non-BALS although
we only have 25 obvious BALs in our sample.

\section{Acknowledgements}

SF would like to thank the University of Hertfordshire for support
whilst this work was being undertaken. 
MJJ acknowledges the support of an RCUK fellowship.

Funding for the SDSS and SDSS-II has been provided by the Alfred
P. Sloan Foundation, the Participating Institutions, the National
Science Foundation, the U.S. Department of Energy, the National
Aeronautics and Space Administration, the Japanese Monbukagakusho, the
Max Planck Society, and the Higher Education Funding Council for
England. The SDSS Web Site is http://www.sdss.org/. 

The SDSS is managed by the Astrophysical Research Consortium for the
Participating Institutions. The Participating Institutions are the
American Museum of Natural History, Astrophysical Institute Potsdam,
University of Basel, University of Cambridge, Case Western Reserve
University, University of Chicago, Drexel University, Fermilab, the
Institute for Advanced Study, the Japan Participation Group, Johns
Hopkins University, the Joint Institute for Nuclear Astrophysics, the
Kavli Institute for Particle Astrophysics and Cosmology, the Korean
Scientist Group, the Chinese Academy of Sciences (LAMOST), Los Alamos
National Laboratory, the Max-Planck-Institute for Astronomy (MPIA),
the Max-Planck-Institute for Astrophysics (MPA), New Mexico State
University, Ohio State University, University of Pittsburgh,
University of Portsmouth, Princeton University, the United States
Naval Observatory, and the University of Washington.

\bibliography{bib}

\begin{thebibliography}{}

\bibitem[\protect\citeauthoryear{{Adelman-McCarthy}, et~al.}{2007}]{ald07}
{Adelman-McCarthy} J.~K., et~al.,  2007, ApJS, 172, 634

\bibitem[\protect\citeauthoryear{{Antonucci}}{{Antonucci}}{1993}]{ant93}
{Antonucci} R.,  1993, ARA\&A, 31, 473

\bibitem[\protect\citeauthoryear{{Baldwin}, {Phillips} \&
  {Terlevich}}{{Baldwin} et~al.}{1981}]{bpt81}
{Baldwin} J.~A.,  {Phillips} M.~M.,    {Terlevich} R.,  1981, PASP, 93, 5

\bibitem[\protect\citeauthoryear{{Becker}, {White}, {Gregg}, {Brotherton},
  {Laurent-Muehleisen} \& {Arav}}{{Becker} et~al.}{2000}]{bec00}
{Becker} R.~H.,  {White} R.~L.,  {Gregg} M.~D.,  {Brotherton} M.~S.,
  {Laurent-Muehleisen} S.~A.,    {Arav} N.,  2000, ApJ, 538, 72

\bibitem[\protect\citeauthoryear{{Becker}, {White} \& {Helfand}}{{Becker}
  et~al.}{1995}]{bwh95}
{Becker} R.~H.,  {White} R.~L.,    {Helfand} D.~J.,  1995, ApJ, 450, 559

\bibitem[\protect\citeauthoryear{{Bennert}, {Jungwiert}, {Komossa}, {Haas} \&
  {Chini}}{{Bennert} et~al.}{2006a}]{ben06s1}
{Bennert} N.,  {Jungwiert} B.,  {Komossa} S.,  {Haas} M.,    {Chini} R.,
  2006a, A\&A, 459, 55

\bibitem[\protect\citeauthoryear{{Bennert}, {Jungwiert}, {Komossa}, {Haas} \&
  {Chini}}{{Bennert} et~al.}{2006b}]{ben06s2}
{Bennert} N.,  {Jungwiert} B.,  {Komossa} S.,  {Haas} M.,    {Chini} R.,
  2006b, A\&A, 456, 953

\bibitem[\protect\citeauthoryear{{Bentz}, et~al.}{2010}]{ben10}
{Bentz} M.~C., et~al.,  2010, ApJ, 716, 993

\bibitem[\protect\citeauthoryear{{Brotherton}}{{Brotherton}}{1996}]{bro96}
{Brotherton} M.~S.,  1996, ApJS, 102, 1

\bibitem[\protect\citeauthoryear{{Cohen}, {Ogle}, {Tran}, {Vermeulen},
  {Miller}, {Goodrich} \& {Martel}}{{Cohen} et~al.}{1995}]{coh95}
{Cohen} M.~H.,  {Ogle} P.~M.,  {Tran} H.~D.,  {Vermeulen} R.~C.,  {Miller}
  J.~S.,  {Goodrich} R.~W.,    {Martel} A.~R.,  1995, ApJL, 448, L77+

\bibitem[\protect\citeauthoryear{{Cohen}}{{Cohen}}{1983}]{coh83}
{Cohen} R.~D.,  1983, ApJ, 273, 489

\bibitem[\protect\citeauthoryear{{Condon}, {Cotton}, {Greisen}, {Yin},
  {Perley}, {Taylor} \& {Broderick}}{{Condon} et~al.}{1998}]{con98}
{Condon} J.~J.,  {Cotton} W.~D.,  {Greisen} E.~W.,  {Yin} Q.~F.,  {Perley}
  R.~A.,  {Taylor} G.~B.,    {Broderick} J.~J.,  1998, AJ, 115, 1693

\bibitem[\protect\citeauthoryear{{Corbett}, et~al.}{2003}]{corb03}
{Corbett} E.~A., et~al.,  2003, MNRAS,
  343, 705

\bibitem[\protect\citeauthoryear{{Decarli}, {Labita}, {Treves} \&
  {Falomo}}{{Decarli} et~al.}{2008}]{dec08}
{Decarli} R.,  {Labita} M.,  {Treves} A.,    {Falomo} R.,  2008, MNRAS, 387,
  1237

\bibitem[\protect\citeauthoryear{{Denney}, et~al.}{2010}]{den10}
{Denney} K.~D., et~al.,  2010, ArXiv e-prints

\bibitem[\protect\citeauthoryear{{Doi}, et~al.}{2009}]{doi09}
{Doi} A., et~al.,  2009, ArXiv e-prints

\bibitem[\protect\citeauthoryear{{Fine}, {Croom}, {Bland-Hawthorn}, {Pimbblet},
  {Ross}, {Schneider} \& {Shanks}}{{Fine} et~al.}{2010}]{me3}
{Fine} S.,  {Croom} S.~M.,  {Bland-Hawthorn} J.,  {Pimbblet} K.~A.,  {Ross}
  N.~P.,  {Schneider} D.~P.,    {Shanks} T.,  2010, ArXiv e-prints

\bibitem[\protect\citeauthoryear{{Fine}, et~al.}{2008}]{me2}
{Fine} S., et~al.,  2008, MNRAS, 390, 1413

\bibitem[\protect\citeauthoryear{{Fukugita}, {Ichikawa}, {Gunn}, {Doi},
  {Shimasaku} \& {Schneider}}{{Fukugita} et~al.}{1996}]{fuk96}
{Fukugita} M.,  {Ichikawa} T.,  {Gunn} J.~E.,  {Doi} M.,  {Shimasaku} K.,
  {Schneider} D.~P.,  1996, AJ, 111, 1748

\bibitem[\protect\citeauthoryear{{Goodrich} \& {Miller}}{{Goodrich} \&
  {Miller}}{1995}]{g+m95}
{Goodrich} R.~W.,  {Miller} J.~S.,  1995, ApJL, 448, L73+

\bibitem[\protect\citeauthoryear{{Gunn}, et~al.}{2006}]{gun06}
{Gunn} J.~E., et~al.,  2006, AJ, 131, 2332

\bibitem[\protect\citeauthoryear{{Heckman}, {Kauffmann}, {Brinchmann},
  {Charlot}, {Tremonti} \& {White}}{{Heckman} et~al.}{2004}]{heck04}
{Heckman} T.~M.,  {Kauffmann} G.,  {Brinchmann} J.,  {Charlot} S.,  {Tremonti}
  C.,    {White} S.~D.~M.,  2004, ApJ, 613, 109

\bibitem[\protect\citeauthoryear{{Jackson} \& {Wall}}{{Jackson} \&
  {Wall}}{1999}]{j+w99}
{Jackson} C.~A.,  {Wall} J.~V.,  1999, MNRAS, 304, 160

\bibitem[\protect\citeauthoryear{{Jarvis} \& {McLure}}{{Jarvis} \&
  {McLure}}{2002}]{j+m02}
{Jarvis} M.~J.,  {McLure} R.~J.,  2002, MNRAS, 336, L38

\bibitem[\protect\citeauthoryear{{Jarvis} \& {McLure}}{{Jarvis} \&
  {McLure}}{2006}]{j+m06}
{Jarvis} M.~J.,  {McLure} R.~J.,  2006, MNRAS, 369, 182

\bibitem[\protect\citeauthoryear{{Kollatschny} \& {Dietrich}}{{Kollatschny} \&
  {Dietrich}}{1996}]{koll96}
{Kollatschny} W.,  {Dietrich} M.,  1996, A\&A, 314, 43

\bibitem[\protect\citeauthoryear{{Kraemer}, {Trippe}, {Crenshaw},
  {Mel{\'e}ndez}, {Schmitt} \& {Fischer}}{{Kraemer} et~al.}{2009}]{kra09}
{Kraemer} S.~B.,  {Trippe} M.~L.,  {Crenshaw} D.~M.,  {Mel{\'e}ndez} M.,
  {Schmitt} H.~R.,    {Fischer} T.~C.,  2009, ApJ, 698, 106

\bibitem[\protect\citeauthoryear{{Labita}, {Treves}, {Falomo} \&
  {Uslenghi}}{{Labita} et~al.}{2006}]{lab06}
{Labita} M.,  {Treves} A.,  {Falomo} R.,    {Uslenghi} M.,  2006, MNRAS, 373,
  551

\bibitem[\protect\citeauthoryear{{Lamastra}, {Bianchi}, {Matt}, {Perola},
  {Barcons} \& {Carrera}}{{Lamastra} et~al.}{2009}]{lam09}
{Lamastra} A.,  {Bianchi} S.,  {Matt} G.,  {Perola} G.~C.,  {Barcons} X.,
  {Carrera} F.~J.,  2009, A\&A, 504, 73

\bibitem[\protect\citeauthoryear{{Magorrian}, et~al.}{1998}]{mag98}
{Magorrian} J., et~al.,  1998, AJ, 115, 2285

\bibitem[\protect\citeauthoryear{{McLure} \& {Dunlop}}{{McLure} \&
  {Dunlop}}{2002}]{m+d02}
{McLure} R.~J.,  {Dunlop} J.~S.,  2002, MNRAS, 331, 795

\bibitem[\protect\citeauthoryear{{Murayama} \& {Taniguchi}}{{Murayama} \&
  {Taniguchi}}{1998a}]{m+t98dc}
{Murayama} T.,  {Taniguchi} Y.,  1998a, ApJL, 503, L115+

\bibitem[\protect\citeauthoryear{{Murayama} \& {Taniguchi}}{{Murayama} \&
  {Taniguchi}}{1998b}]{m+t98clr}
{Murayama} T.,  {Taniguchi} Y.,  1998b, ApJL, 497, L9+

\bibitem[\protect\citeauthoryear{{Nagao}, {Murayama} \& {Taniguchi}}{{Nagao}
  et~al.}{2001}]{nag01}
{Nagao} T.,  {Murayama} T.,    {Taniguchi} Y.,  2001, PASJ, 53, 629

\bibitem[\protect\citeauthoryear{{Onken}, {Ferrarese}, {Merritt}, {Peterson},
  {Pogge}, {Vestergaard} \& {Wandel}}{{Onken} et~al.}{2004}]{onk04}
{Onken} C.~A.,  {Ferrarese} L.,  {Merritt} D.,  {Peterson} B.~M.,  {Pogge}
  R.~W.,  {Vestergaard} M.,    {Wandel} A.,  2004, ApJ, 615, 645

\bibitem[\protect\citeauthoryear{{Orr} \& {Browne}}{{Orr} \&
  {Browne}}{1982}]{o+b82}
{Orr} M.~J.~L.,  {Browne} I.~W.~A.,  1982, MNRAS, 200, 1067

\bibitem[\protect\citeauthoryear{{Osterbrock}}{{Osterbrock}}{1977}]{ost77}
{Osterbrock} D.~E.,  1977, ApJ, 215, 733

\bibitem[\protect\citeauthoryear{{Peterson} \& {Wandel}}{{Peterson} \&
  {Wandel}}{1999}]{p+w99}
{Peterson} B.~M.,  {Wandel} A.,  1999, ApJL, 521, L95

\bibitem[\protect\citeauthoryear{{Rengelink}, {Tang}, {de Bruyn}, {Miley},
  {Bremer}, {Roettgering} \& {Bremer}}{{Rengelink} et~al.}{1997}]{ren97}
{Rengelink} R.~B.,  {Tang} Y.,  {de Bruyn} A.~G.,  {Miley} G.~K.,  {Bremer}
  M.~N.,  {Roettgering} H.~J.~A.,    {Bremer} M.~A.~R.,  1997, A\&AS, 124, 259

\bibitem[\protect\citeauthoryear{{Reyes}, et~al.}{2008}]{rey08}
{Reyes} R., et~al.,  2008,
  AJ, 136, 2373

\bibitem[\protect\citeauthoryear{{Richards}, et~al.}{2002}]{rich02b}
{Richards} G.~T., et~al.,  2002, AJ, 123, 2945

\bibitem[\protect\citeauthoryear{{Richards}, et~al.}{2002}]{rich02}
{Richards} G.~T., et~al.,  2002,
  AJ, 124, 1

\bibitem[\protect\citeauthoryear{{Risaliti}, {Elvis}, {Bianchi} \&
  {Matt}}{{Risaliti} et~al.}{2010}]{ris10}
{Risaliti} G.,  {Elvis} M.,  {Bianchi} S.,    {Matt} G.,  2010, MNRAS, 406, L20

\bibitem[\protect\citeauthoryear{{Schmitt}}{{Schmitt}}{1998}]{sch98}
{Schmitt} H.~R.,  1998, ApJ, 506, 647

\bibitem[\protect\citeauthoryear{{Schneider}, et~al.}{2007}]{sch07}
{Schneider} D.~P., et~al.,  2007,
  AJ, 134, 102

\bibitem[\protect\citeauthoryear{{Shuder} \& {Osterbrock}}{{Shuder} \&
  {Osterbrock}}{1981}]{s+o81}
{Shuder} J.~M.,  {Osterbrock} D.~E.,  1981, ApJ, 250, 55

\bibitem[\protect\citeauthoryear{{Simpson}}{{Simpson}}{2005}]{sim05}
{Simpson} C.,  2005, MNRAS, 360, 565

\bibitem[\protect\citeauthoryear{{Stoughton}, et~al.}{2002}]{sto02}
{Stoughton} C., et~al.,  2002, AJ, 123, 485

\bibitem[\protect\citeauthoryear{{Trump}, et~al.}{2006}]{tru06}
{Trump} J.~R., et~al.,  2006, ApJS, 165,
  1

\bibitem[\protect\citeauthoryear{{Urry} \& {Padovani}}{{Urry} \&
  {Padovani}}{1995}]{u+p94}
{Urry} C.~M.,  {Padovani} P.,  1995, PASP, 107, 803

\bibitem[\protect\citeauthoryear{{Vestergaard}, {Wilkes} \&
  {Barthel}}{{Vestergaard} et~al.}{2000}]{vwb00}
{Vestergaard} M.,  {Wilkes} B.~J.,    {Barthel} P.~D.,  2000, ApJL, 538, L103

\bibitem[\protect\citeauthoryear{{Weymann}, {Morris}, {Foltz} \&
  {Hewett}}{{Weymann} et~al.}{1991}]{wey91}
{Weymann} R.~J.,  {Morris} S.~L.,  {Foltz} C.~B.,    {Hewett} P.~C.,  1991,
  ApJ, 373, 23

\bibitem[\protect\citeauthoryear{{Willott}, {Rawlings}, {Blundell} \&
  {Lacy}}{{Willott} et~al.}{2000}]{wil00}
{Willott} C.~J.,  {Rawlings} S.,  {Blundell} K.~M.,    {Lacy} M.,  2000, MNRAS,
  316, 449

\bibitem[\protect\citeauthoryear{{Wills} \& {Browne}}{{Wills} \&
  {Browne}}{1986}]{w+b86}
{Wills} B.~J.,  {Browne} I.~W.~A.,  1986, ApJ, 302, 56

\bibitem[\protect\citeauthoryear{{Wilman}, et~al.}{2008}]{wilm08}
{Wilman} R.~J., et~al.,  2008, MNRAS, 388, 1335

\bibitem[\protect\citeauthoryear{{York}, et~al.}{2000}]{yor00}
{York} D.~G., et~al.,  2000,
  AJ, 120, 1579

\bibitem[\protect\citeauthoryear{{Zhang}, {Wang}, {Dong} \& {Lu}}{{Zhang}
  et~al.}{2008}]{zha08}
{Zhang} K.,  {Wang} T.,  {Dong} X.,    {Lu} H.,  2008, ApJL, 685, L109

\bibitem[\protect\citeauthoryear{{Zhou}, {Wang}, {Wang}, {Wang}, {Yuan} \&
  {Lu}}{{Zhou} et~al.}{2006}]{zho06}
{Zhou} H.,  {Wang} T.,  {Wang} H.,  {Wang} J.,  {Yuan} W.,    {Lu} Y.,  2006,
  ApJ, 639, 716

\end{thebibliography}

\end{document}